\documentclass{article}
\usepackage[margin=1in]{geometry}
\usepackage{graphicx}
\usepackage{url}
\usepackage{titlesec}
\usepackage{pifont}
\newcommand{\cmark}{\ding{51}}
\newcommand{\xmark}{\ding{55}}

\usepackage[pdfauthor={Arin Upadhyay}, pdftitle={EFPIX: An Encrypted Flood Protocol for Metadata-Resistant Communication}, pdfkeywords={encryption, flooding, anonymity, protocol, decentralized, cryptography, mesh, relay, metadata, privacy}]{hyperref}

\title{\textbf{EFPIX: An Encrypted Flood Protocol for Metadata-Resistant Communication}}
\author{Arin Upadhyay \\ \texttt{arinupadhyay.cs@gmail.com}}
\date{\today}
\begin{document}
\maketitle

\begin{abstract}
We propose EFPIX, a flood-based relay protocol for encrypted, metadata-resistant, and spam-tolerant communication. By default, messages are end-to-end encrypted, and metadata is hidden from relaying nodes, while proof-of-work, deduplication, and aging deter spam and replay without requiring accounts or trusted authorities. The protocol is topology-agnostic, resilient to infrastructure failures, and requires no central server. It is designed for asynchronous privacy-critical communication, where anonymity and resilience matter more than throughput.
\end{abstract}

\section{Introduction}
In recent years, governments and invasive intelligence agencies have transformed the Internet to erode privacy rights using various surveillance programs, laws, and backdoors. \cite{bamford2012} \cite{heilman2014} \cite{pegasus2023} They are not hesitant to restrict or ban encryption \cite{apple2025}, and have the power to subpoena any data, track network traffic, and force shutdown of services that counter these issues. This has made activism and whistle-blowing extremely dangerous. \cite{snowden2014} It is difficult to securely communicate, even if one's own devices are not compromised. \\
An attractive solution to this problem is an encrypted flood protocol. Such protocols provide secure peer-to-peer communication while also being topology-agnostic and resilient to infrastructure failures or takedowns. This has many possible applications, including journalism where anonymity is of utmost importance or military fields, where the existence and maintenance of a central server is not feasible. They are also suitable for broadcast-type messages that reach all nodes in the network, such as emergency rescue calls, disaster warnings, news distribution, etc. \\
EFPIX (Encrypted Flood Protocol for Information eXchange) is designed with the above issues in mind. The protocol is designed to conceal the data and prevent identification of the source or destination of the packet while offering protection against network-level spam and abuse. It is possible to build higher-layer protocols onto this protocol or compatible modifications to serve various use cases. \\
EFPIX is not intended for high-bandwidth or real-time applications such as live video streaming or gaming. Instead, it is optimized for asynchronous, privacy-critical communication, for example, email, secure file drops, whistleblower reports, offline messaging, or distributed alerts, where anonymity and resilience outweigh performance.

\section{Terms \& Concepts}
\texttt{Alias}: A lookup key into the Contact Map. To message a given receiver, the sender retrieves that receiver's public key and the sender's own alias for them from the local contact map, but only the alias is included in the packet, while the public key is used to encrypt it. The receiver uses the received alias to look up the sender's public key in their own contact map. It is agreed upon before communication begins and can differ per correspondent. \\
\texttt{Contact Map}: A local table maps (receiver/their alias) to {public Key, sender/my alias}, established before communication begins. To prevent adversaries from linking packets to a common origin, users should practice good operational security by using unique aliases and key-pairs for each correspondent. The management of these identities is delegated to the user or higher-level applications, similar to identity management in protocols like Bitmessage. In practice, key and alias exchange may itself be performed over EFPIX at a higher layer as a "Connection Request". \\
\texttt{PoW Nonce}: A proof-of-work (PoW) requirement must be met to minimize spam and tampering similar to Bitcoin \cite{bitcoin2008}. However, nonce is excluded from the deduplication hash calculation. This is done because an attacker can find a new valid nonce, resulting in a different deduplication hash even though the packet is the same. The PoW difficulty must be calibrated against the aggregate cost of trial decryption imposed on relaying neighbors, the amount of spam a given deployment is willing to tolerate in exchange for lower latency, and the generation-time constraints of legitimate senders. \\
\texttt{Deduplication}: Deduplication hash is calculated by a relaying node on the packet, excluding the PoW nonce. It is used to determine whether the node has already seen this packet, which prevents infinite looping relays. Every node keeps a record of hashes of previously relayed packets.\\
\texttt{Timestamp}: A compactly encoded representation of the time of the packet's creation. \\
\texttt{Max Age}: It is the difference between timestamp of reception and timestamp of creation. The maximum packet age should be less than how long the hash is stored by the node and can be different for different nodes.

\section{Encoding}
Before encryption, the desired message is concatenated with the timestamp and the sender alias that is retrieved from the contact map.
This concatenated data is used to generate a signature using the sender’s private key which is then also concatenated with the payload. 
This is encrypted using the receiver's public key which is also retrieved from the local contact map. Finally, PoW nonce and encrypted blob are concatenated to produce the final encoded packet. \\

\begin{figure}[h]
    \centering
    \includegraphics[width=0.7\linewidth]{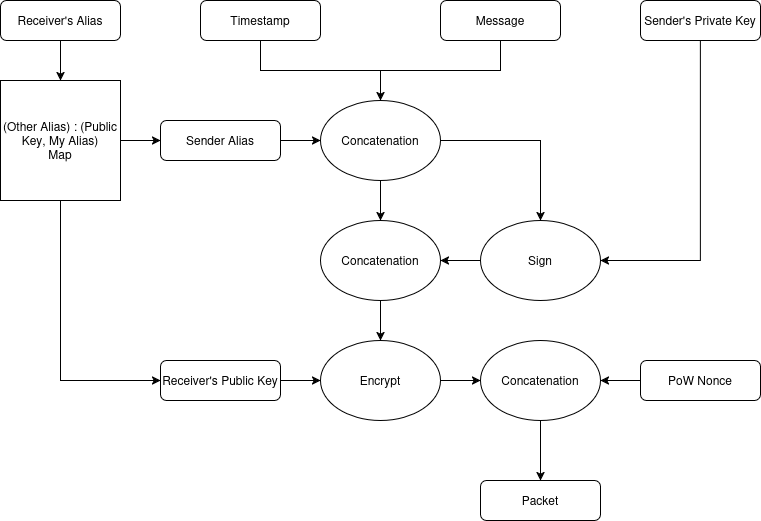}
    \caption{Message encoding process}
\end{figure}

\section{Decoding and Relaying}

PoW is verified and upon failure the packet is discarded. If the node has seen the hash, the packet is discarded. If not, the hash is added to the record and the packet is relayed to all neighbors. Then the node attempts decryption on the encrypted part. If decryption fails, the packet is discarded. Otherwise, the original components are extracted (timestamp of creation, sender alias, message and signature). The node then looks up the sender’s public key using the sender alias. If the alias is not found, the message is treated as "anonymous", and the signature is not verified. Otherwise, the decryption product is used to verify the signature using the sender’s public key. The node verifies that the age of the packet is less than the max age and upon failure the packet is discarded. \\

\begin{figure}[h]
    \centering
    \includegraphics[width=0.9\linewidth]{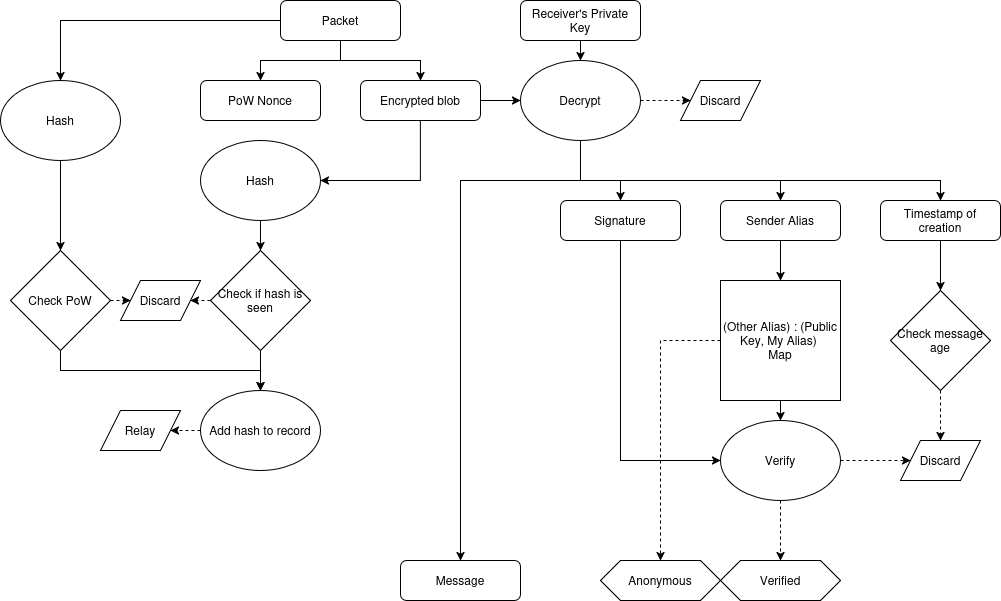}
    \caption{Message decoding and relay process}
\end{figure}

\section{Application-based Improvements}
These improvements are not strictly required but are strongly advised to be included in implementations, like ephemeral keys and recipient binding. \\
\texttt{Ephemeral keys}: Compromising a private key compromises all past messages encrypted to that key. Ephemeral keys or periodic key rotation should be used.\\
\texttt{Recipient Binding}: Reusing an alias across correspondents, could lead to surreptitious forwarding. Including the receiver's alias into the pre-encryption data and signature which the receiver will match with its alias from the contact map can solve this. \\
\texttt{Version \& type header}: A very small header can be added to differentiate versions and packet types. This allows the same network and protocol to support different features at the same time like broadcast and link packets. Every packet, regardless of version or type, must include the PoW nonce field. The PoW difficulty, however, may vary by version or type. \\
\texttt{Broadcast}: A packet type, distinguished via the type header, for one-to-many messages intended to be read by any receiving node rather than a single recipient. Instead of an encrypted blob addressed to a receiver's public key, the message and the sender's alias or public key are included directly. A node verifies the signature against that embedded key rather than looking one up via the sender alias. PoW, deduplication, and aging are applied identically to Broadcast packets as in the core protocol. \\
\texttt{Tagged Decryption}: A short, non-secret tag can be prepended to the encrypted blob to indicate the intended receiver. Nodes can quickly check this tag before attempting decryption. However, this may also increase the risk of deanonymization.\\
\texttt{HMAC-based Signatures}: Instead of public-key signatures, implementations can optionally use keyed HMAC signatures for faster message integrity verification. It can also provide deniability of authorship to the sender. However, this requires a per-sender secret between sender and receiver. \\
\texttt{Link}: A fast but less anonymous abstract channel that can be set up between two nodes to trade features like forward secrecy and metadata protection for speed. It can be set up using the base packet type (described in Sections 3–4 and referred to as a Flood packet, as distinct from Link packets) to negotiate parameters the chosen methods require, such as a decryption tag, symmetric key, or HMAC secret key.\\
\texttt{Dummy traffic}: Attempts of traffic analysis can be nullified by nodes creating and relaying dummy packets and/or relaying packets only after a short duration of random time.\\
\texttt{Better Seen-Hash}: The record of hashes of previously relayed packets can grow uncontrollably. It may be cleared after a random time or data structures like bloom filters may be used.\\
\texttt{Alternative Proof-of-Work Mechanisms}: Standard proof-of-work is vulnerable to adversaries with high-end parallel hardware (GPUs, ASICs). A better alternative is the use of Verifiable Delay Functions (VDFs) \cite{vdf2019}. VDFs require a specific duration of sequential computation, effectively neutralizing the advantage of parallel processors while providing more predictable packet generation times.\\
\texttt{Tokens}: Replay is prevented via deduplication using PoW hash, signatures, and aging. More rules can be implemented on top, including session or one-time tokens or message counters.\\
\texttt{Sector-based routing}: To improve efficiency in large networks, the packet can be flooded in only specific regions called \texttt{sectors}, which is an arbitrary group of nodes. However, this may come at the cost of a greater risk of deanonymization of the receiver and sender. Moreover, the sender must know which sector the receiver is in.\\
\texttt{Time To Live}: TTL is used in structured systems, but here topologies are unknown and dynamic. It may also increase the risk of deanonymization. It could be used for niche cases.\\

EFPIX deliberately prioritizes maximum anonymity and resilience over performance. Optional optimizations such as sector-based routing, TTL, or Link may increase scalability or speed but reduce anonymity. The choice depends entirely on application requirements.

\section{Example Instantiation}
This example uses Curve25519 (X25519 for key exchange, Ed25519 for signatures) \cite{curve255192006}, BLAKE2b for hashing \cite{blake22013}, and authenticated encryption with XChaCha20-Poly1305.\\

Pre-encryption data format (344 bytes): \\
\begin{table}[h]
\centering
\begin{tabular}{|c|c|c|c|}
\hline
\textbf{Timestamp} & \textbf{Sender Alias} & \textbf{Message} & \textbf{Signature} \\
\hline
8 bytes & 16 bytes & 256 bytes & 64 bytes \\
\hline
\end{tabular}
\end{table}

Flood packet layout (424 bytes): \\
\begin{table}[h]
\centering
\begin{tabular}{|c|c|c|c|c|}
\hline
\textbf{PoW Nonce} & \textbf{MAC} & \textbf{Nonce} & \textbf{Ephemeral PK} & \textbf{Ciphertext} \\
\hline
8 bytes & 16 bytes & 24 bytes & 32 bytes & 344 bytes \\
\hline
\end{tabular}
\end{table}

\section{Threat Model and Security Analysis}
We assume endpoint devices and private keys are not compromised and full topological compromise is out of scope. This analysis applies to core protocol behavior only. Optional extensions may alter these properties.

\subsection{Adversary Model}
\texttt{Passive Eavesdropper}: Observes network traffic without modifying it.\\
\texttt{Active Adversary}: Modifies packets and may act as a participating node.\\
\texttt{Compromised Node}: A legitimate node acting maliciously, leaking keys or dropping packets.\\
\texttt{Global Passive Observer (GPO)}: Observes all network traffic, attempting to correlate timing or metadata.\\
\texttt{Sybil Attacker}: Launches many fake nodes to gain influence or disrupt hashing and routing.\\
\texttt{Circling Adversary}: Controls a majority of surrounding nodes to localize packet origin.\\
\texttt{Replay Attacker}: Re-broadcasts valid packets to cause confusion or traffic analysis. \\
\texttt{DoS Attacker}: Floods the network with packets to slow down or overwhelm the network. 

\subsection{Security Properties}

\texttt{Passive Eavesdropping}: Encryption hides packet content and metadata, as an opaque encrypted blob to observers. \\
\texttt{Active Modification}: Packets are protected by digital signatures and hash verification and tampered packets are discarded. \\
\texttt{Compromised Nodes}: Flooding ensures alternate paths unless a compromised node becomes a topological choke point. \\
\texttt{Global Passive Observer (GPO)}: Deduction of the destination of a packet is not possible because traffic will always appear uniform in time since a node only processes a packet after relaying to its neighbors. However, timing-based origin correlation could be possible. This can be deterred by random relay delays and dummy traffic. \\
\texttt{Sybil Attacker}: The protocol is designed so that a node places no trust in any other node or group of nodes. Gaining influence over the network does not by itself affect routing or spam resistance, since PoW and deduplication are verified independently by every node. However, a Sybil swarm positioned to fully encircle a sender, known as a circling attack, can correlate a packet to its sender. EFPIX does not claim to solve this, as it is an inherent consequence of fully flooded mesh networking. \\
\texttt{Replay Attacker}: Timestamps in signatures, PoW hash, and age rules prevent simple replays. A replay will result in an immediate packet drop by the attacker's neighbors since the hash is already seen, and by the receiver if the age is exceeded. Message counters and tokens can be used on top.\\
\texttt{Denial of Service}: PoW or VDFs deter such attacks by being economically and computationally very expensive. 

\section{Comparison with Related Protocols}

\begin{table}[h]
\centering
\begin{tabular}{|c|c|c|c|c|c|}
\hline
\textbf{Feature} & \textbf{EFPIX} & \textbf{Tor \cite{tor2004}} & \textbf{Bitmessage \cite{bitmessage2012}} & \textbf{Reticulum \cite{reticulum2016}} & \textbf{Briar \cite{briar2018}} \\
\hline
Topology-agnostic          & \cmark & \xmark & \cmark & \cmark & \xmark \\
End-to-End Encryption      & \cmark & Partial & \cmark & \cmark & \cmark \\
Metadata Protection        & \cmark & Partial & Partial & Partial & Partial \\
Spam Resistance            & \cmark & N/A & \cmark & \xmark & N/A \\
Broadcast Suitability      & \cmark & \xmark & \cmark & Partial & \xmark \\
No Internet Required       & \cmark & \xmark & \xmark & \cmark & \cmark \\
Low-Bandwidth Link Support & \cmark & \xmark & \xmark & \cmark & \xmark \\
\hline
\end{tabular}
\caption{Comparison of EFPIX with selected secure communication protocols}
\end{table}

\section{Conclusion}
EFPIX is a protocol that provides security of data and identity while protecting communication from infrastructure failure, surveillance and network abuse. It also provides many compatible optional improvements that can be used in different scenarios and applications. It sets itself apart from existing solutions in the field by providing resistance to spam and faults. Its resilient and secure design maximizes delivery and anonymity making it an ideal option for disconnected, totalitarian, disaster-struck areas. \\


\begin{thebibliography}{13}
\bibitem{bamford2012}
J. Bamford, \textit{The NSA Is Building the Country’s Biggest Spy Center (Watch What You Say)}, 
\url{https://www.wired.com/2012/03/ff-nsadatacenter}, 2012.

\bibitem{heilman2014}
E. Heilman, \textit{A Brief History of NSA Backdoors}, 
\url{https://ethanheilman.tumblr.com/post/70646748808/a-brief-history-of-nsa-backdoors}, 2014.

\bibitem{pegasus2023}
European Parliament, \textit{Investigation of the use of Pegasus and
equivalent surveillance spyware}, 
\url{https://www.europarl.europa.eu/RegData/etudes/ATAG/2023/747923/EPRS_ATA(2023)747923_EN.pdf}, 2023.

\bibitem{apple2025}
\textit{Apple pulls data protection tool after UK government security row}, 
\url{https://www.bbc.com/news/articles/cgj54eq4vejo}, 2025.

\bibitem{snowden2014}
\textit{Edward Snowden: Leaks that exposed US spy programme}, 
\url{https://www.bbc.com/news/world-us-canada-23123964}, 2014.

\bibitem{bitcoin2008}
S. Nakamoto, \textit{Bitcoin: A Peer-to-Peer Electronic Cash System},  \url{https://bitcoin.org/bitcoin.pdf}, 2008.

\bibitem{vdf2019}
Dan Boneh, Joseph Bonneau, Benedikt Bunz, and Ben Fisch, \textit{Verifiable Delay Functions},  
\url{https://eprint.iacr.org/2018/601.pdf}, 2019.

\bibitem{curve255192006}
Daniel J. Bernstein, \textit{Curve25519: new Diffie-Hellman speed records},  
\url{https://cr.yp.to/ecdh/curve25519-20060209.pdf}, 2005.

\bibitem{blake22013}
Jean-Philippe Aumasson, Samuel Neves, Zooko Wilcox-O’Hearn, and Christian Winnerlein, \textit{BLAKE2: simpler, smaller, fast as MD5},  
\url{https://www.blake2.net/blake2.pdf}, 2013.

\bibitem{tor2004}
R. Dingledine, N. Mathewson, and P. Syverson, \textit{Tor: The Second-Generation Onion Router}, \url{https://www.usenix.org/events/sec04/tech/dingledine.html}, 2004.

\bibitem{bitmessage2012}
J. Warren, \textit{Bitmessage: A Peer-to-Peer Message Authentication and Delivery System}, \url{https://bitmessage.org/bitmessage.pdf}, 2012.

\bibitem{reticulum2016}
M. Qvist, \textit{Reticulum Network Stack}, \url{https://reticulum.network/manual/index.html}, 2016.

\bibitem{briar2018}
Briar Project, \textit{Briar: How it works} \url{https://briarproject.org/how-it-works/}, 2018.

\end{thebibliography}
\end{document}